\documentclass[a4paper,twocolumn,11pt,accepted=2026-05-14]{quantumarticle}
\pdfoutput=1
\usepackage{amsmath}
\usepackage{amssymb}
\usepackage{booktabs}
\usepackage[allcolors=blue]{hyperref}
\usepackage[english]{babel}
\usepackage[numbers,sort]{natbib}
\usepackage{array}
\usepackage[english]{babel}
\usepackage{orcidlink}
\usepackage[utf8]{inputenc}
\usepackage[T1]{fontenc}
\usepackage{graphicx}
\usepackage{physics}
\usepackage{xcolor}

\usepackage[normalem]{ulem}

\newcommand{\void}[1]{}

\begin{document}

\title{Hidden time-nonlocal Floquet symmetries}

\author{Sigmund Kohler\,\orcidlink{0000-0003-3668-8030}}
\affiliation{Quantum Advanced Research Center (QuARC), CSIC, Sor Juana Inés de la Cruz 3, 28049 Madrid, Spain}
\affiliation{Instituto de Ciencia de Materiales de Madrid (ICMM), CSIC, Sor Juana Inés de la Cruz 3, 28049 Madrid, Spain}
\email{sigmund.kohler@csic.es}

\author{Jes\'us Casado-Pascual\,\orcidlink{0000-0003-4118-5809}}
\affiliation{F{\'\i}sica  Te\'orica,  Universidad  de  Sevilla,  Apartado  de
Correos 1065, E-41080 Sevilla, Spain}
\affiliation{Multidisciplinary Unit for Energy Science, Universidad de Sevilla, E-41080 Sevilla, Spain}
\email{jcasado@us.es}

\begin{abstract}
We investigate the Floquet spectrum of a detuned, driven two-level system and show that it
exhibits exact quasienergy crossings when the detuning is an integer multiple of the energy 
quantum of the driving field. This behavior can be explained by a hidden time-nonlocal 
parity, which allows the Floquet modes to be classified as even or odd. Then a 
generic feature is the emergence of exact crossings between quasienergies of different 
parity. A constructive proof of the existence of the symmetry is based on a scalar 
recurrence relation. Moreover, we present a general scheme for its numerical computation, 
which can be applied to models beyond the two-level system. Analytical results are 
illustrated with numerical data.
\end{abstract}

\maketitle

\section{Introduction}
\label{sec:intro}

Quasienergies reflect the spectral properties of ac-driven quantum systems
and are a cornerstone of Floquet theory~\cite{ShirleyPR65, SambePRA73,
Hanggi98}.  They determine the phase factors of the long-time dynamics and,
thus, their splittings set the corresponding time scales.  Of
particular interest are quasienergy crossings at which these time scales
diverge so that the quantum dynamics may become frozen.  A prominent effect
that relies on this is coherent destruction of tunneling (CDT) \cite{GrossmannPRL91},
which can already be understood within a two-level
approximation~\cite{GrossmannEL92}.  This appealing prediction has spurred
a wealth of experiments with double quantum dots \cite{StehlikPRB12,
ForsterPRL14}, superconducting qubits \cite{SillanpaaPRL06, BernsNL08},
atoms in optical lattices \cite{LignierPRL07}, and optical waveguides \cite{DellaVallePRL07}.
Lately, exact quasienergy crossings attracted attention, because in their
vicinity, the dissipative behavior is rather sensitive to small parameter
variation~\cite{EngelhardtPRL19, KohlerPRA24}.

Quasienergies are eigenvalues of the Floquet Hamiltonian, which is a
Hermitian operator in Sambe space, i.e., in
the product space of the underlying Hilbert space and that of time-periodic
functions~\cite{ShirleyPR65, SambePRA73}.  As such, they exhibit generic
features of quantum-mechanical spectra, in particular level repulsion~\cite{Haake2018}.
Therefore, as a function of any system parameter,
quasienergies are expected to form avoided crossings (also called
anti-crossings) unless a symmetry or integral of motion is present.
For the example of CDT in an undetuned two-level
system, the emergence of exact crossings is enabled by a spatio-temporal
symmetry known as generalized parity~\cite{PeresPRL91}.

For large driving frequencies, the CDT Hamiltonian can be approximated by a
time-independent effective Hamiltonian in which the tunnel matrix element
is renormalized by the zeroth-order Bessel function of the first kind.  At
its roots, tunneling is suppressed~\cite{GrossmannEL92}.  This
approximation scheme can be generalized to the presence of an ``integer
detuning,'' i.e., a detuning that matches $n$ energy quanta of the driving
field.  The resulting renormalization by the $n$th-order Bessel function has
been verified numerically~\cite{StrassPRL05, AshhabPRA07, IvakhnenkoPR23}
and observed experimentally~\cite{StehlikPRB12, ForsterPRL14}.  Since all
Bessel functions of the first kind possess roots, within a high-frequency
approximation one finds exact crossings.  This raises the question of whether,
beyond the approximation, these crossings are indeed exact or just narrowly
avoided.  And if they are exact, to which symmetry or integral of motion
can they be attributed?  Related questions have been addressed recently also
for the (time-independent) Rabi Hamiltonian~\cite{AshhabPRA20,
MangazeevJPA21}, for which exact level crossings are enabled by a 
hidden symmetry~\cite{ReyesBustosJPA21}.

Here, we develop an approach for finding hidden time-nonlocal symmetries of
Floquet systems and use it to explain the emergence of exact crossings in
the Floquet spectrum of the driven two-level system.  The work is organized
as follows.  In Sec.~\ref{sec:model}, we introduce our model and, 
from numerical findings, conjecture the condition for the emergence of exact
quasienergy crossings, and consequently the existence of a hidden symmetry.  
The strategy for finding such symmetry and its application to our model is presented in
Sec.~\ref{sec:LRI}.  In Sec.~\ref{sec:numerics}, we develop a scheme for
its numerical computation, while conclusions are drawn in
Sec.~\ref{sec:conclusions}.  An intuitive derivation for the particular
case $n=1$ and details of the numerical approach are given in the Appendix.

\section{The driven two-level system}
\label{sec:model}

We consider the driven two-level system described by the pseudospin Hamiltonian
\begin{equation}
H(t) = \frac{\epsilon}{2}\sigma_z
 + \beta\sigma_x + \alpha\sigma_z\cos(\Omega t),
\label{H}
\end{equation}
with detuning $\epsilon$, tunneling matrix element $\beta$, driving amplitude
$\alpha$, and frequency $\Omega$.  For convenience and without loss of
generality, we assume $\alpha,\,\beta,\,\epsilon,\,\Omega\geq 0$ and choose units
with $\hbar=1$.

For such a time-periodic Hamiltonian, the Floquet theorem states that a
complete set of solutions of the Schr\"odinger equation is of the form
$\ket{\psi(t)} = e^{-iqt} \ket{\phi(t)}$, with quasienergy $q$.  The
Floquet mode $\ket{\phi(t)} = \ket{\phi(t+T)}$, with $T=2\pi/\Omega$,
shares the time-periodicity of the Hamiltonian.  Therefore, it can be
considered an
element of Sambe space, i.e., Hilbert space extended by the space of $T$-periodic
functions~\cite{ShirleyPR65, SambePRA73}.  By inserting this ansatz into the
Schr\"odinger equation one readily finds that the Floquet modes obey the
eigenvalue equation
\begin{equation}
\Big( H(t) - i\frac{\partial}{\partial t}\Big) \ket{\phi(t)}
= q\ket{\phi(t)},
\label{Feq}
\end{equation}
with the Floquet Hamiltonian, or quasienergy operator, $H(t)-i\partial_t$.
It is straightforward to show that when $\ket{\phi(t)}$ is a Floquet
mode with quasienergy $q$, then, for any integer $k$, $e^{ik\Omega t}\ket{\phi(t)}$
is a Floquet mode with quasienergy $q+k\Omega$.  Both modes are equivalent as they 
correspond to the same solution of the Schr\"odinger equation.  
This constitutes the Brillouin-zone structure of the Floquet spectrum. 
Irrespective of the choice of the Brillouin zone, a complete set of non-equivalent 
Floquet modes at equal times forms an orthonormal basis of the Hilbert space~\cite{SambePRA73}.

\begin{figure}
\centerline{\includegraphics{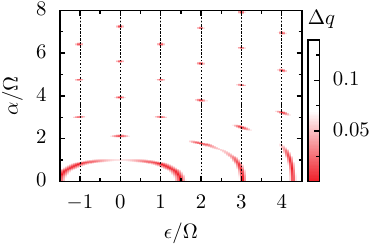}}
\caption{Minimal quasienergy splitting as a function of the detuning and
the driving amplitude for tunnel coupling $\beta=1.3\,\Omega$.  The color
scale is chosen such that regions with particularly small values are highlighted.
The vertical dashed lines mark detunings that are integer multiples of
the driving frequency, $\epsilon = n\Omega$.}
\label{fig:splitting}
\end{figure}

\begin{figure}
\centerline{\includegraphics{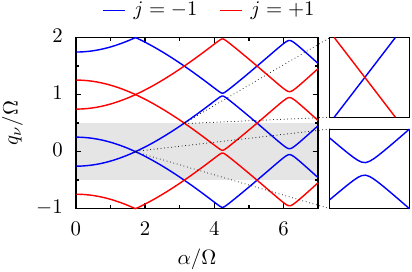}}
\caption{Floquet spectrum of the driven two-level system as a function of the
driving amplitude $\alpha$ for tunneling matrix element $\beta=2.7\,\Omega$ and
detuning $\epsilon=\Omega$ (i.e., $n=1$). The gray box marks the first
Brillouin zone.  The parity $j=\pm 1$, derived below, is indicated by the
color---red for $+1$ and blue for $-1$. Equivalent modes in neighboring Brillouin zones have opposite
parity.  The zooms indicate that only quasienergies with the same
parity form avoided crossings, while those with different parity cross
exactly.}
\label{fig:spectrum}
\end{figure}

Figure~\ref{fig:splitting} shows the minimal quasienergy splitting of
Hamiltonian \eqref{H} as a function of detuning and driving amplitude.  Unless the
driving amplitude is rather small, the regions with tiny splittings are
centered at integer values of $\epsilon/\Omega$.  Below, we demonstrate that
for such integer detuning the splittings are indeed zero.  This implies that
then the Floquet spectra as a function of the driving amplitude exhibit
exact crossings.  In Fig.~\ref{fig:spectrum}, this is illustrated for the
special case $\epsilon = \Omega$.  Therefore, in accordance with the generic
properties of quantum-mechanical spectra~\cite{Haake2018}, we can conclude
that the Hamiltonian~\eqref{H} must possess a symmetry or an integral of
motion that characterizes the Floquet modes.

In the absence of detuning ($\epsilon=0$), the relevant symmetry is the well-known generalized parity $G = \sigma_x P$, where
$P=\exp[(T/2)\partial_t]$ shifts time by half a driving period~\cite{PeresPRL91, GrossmannEL92}. The Floquet modes are eigenstates of $G$, i.e., $G\ket{\phi_\nu(t)} = \sigma_x \ket{\phi_\nu(t+T/2)}=j_\nu \ket{\phi_\nu(t)}$.  Owing to the $T$-periodicity
of the Floquet modes, $G^2\ket{\phi_\nu(t)} = \ket{\phi_\nu(t)}$,
which implies that the eigenvalues $j_\nu$ are $\pm 1$.

\section{Time-nonlocal symmetries}
\label{sec:LRI}

For nonzero detuning $\epsilon\neq0$, Hamiltonian~\eqref{H} lacks an obvious symmetry.
Therefore, we search for a hidden spatio-temporal symmetry with the structure of the generalized
parity~\cite{PeresPRL91}. Specifically, we look for a generally time-dependent operator $Q(t)$
such that $J(t) = Q(t)P$ acts as a parity-like symmetry operator on the Floquet modes 
$\ket{\phi_\nu(t)}$, i.e.,
\begin{equation}
J(t)\ket{\phi_\nu(t)} = j_\nu \ket{\phi_\nu(t)},
\label{symmJ}
\end{equation}
with $j_\nu = \pm 1$. The corresponding relation for $Q(t)$ reads
\begin{equation}
Q(t)\ket{\phi_\nu(t+T/2)} = j_\nu \ket{\phi_\nu(t)} .
\label{symmQ}
\end{equation}
Since two equivalent Floquet modes from neighboring Brillouin zones differ by a
phase factor $e^{i\Omega t}$, they will have opposite parity.
This is already the case for the generalized parity in the absence of detuning 
\cite{PeresPRL91}.  It underlines that while equivalent modes correspond to the same 
solution of the Schr\"odinger equation, they are different solutions of the eigenvalue
equation~\eqref{Feq}.

From the $T$-periodicity of the Floquet modes follows 
$Q(t+T/2)\ket{\phi_\nu(t)} = j_\nu \ket{\phi_\nu(t+T/2)}$.
Therefore, $Q(t)Q(t+T/2)\ket{\phi_\nu(t)} = \ket{\phi_\nu(t)}$ for all Floquet modes
and hence $Q(t)Q(t+T/2) = \mathbb{I}$. Moreover, since $|j_\nu|=1$, Eq.~\eqref{symmQ}
implies that $Q(t)$ maps an orthonormal basis into another orthonormal basis, and 
therefore it must be unitary, which leads to
\begin{equation}
Q^\dagger(t) = Q(t+T/2).
\label{Qd}
\end{equation}
Using this relation twice yields that $Q(t)$ must be a $T$-periodic function of time.

To compute $Q(t)$ for a given Hamiltonian, we derive its equation of motion by applying $i\partial_t$
to Eq.~\eqref{symmQ}. The resulting time derivatives of the Floquet modes can be replaced using the
Floquet equation~\eqref{Feq}, yielding 
$i \partial_t{Q}(t) \ket{\phi_\nu(t+T/2)} = [H(t)Q(t) - Q(t)H(t+T/2)] \ket{\phi_\nu(t+T/2)}$. 
Since this equation holds for all Floquet modes, it follows that
\begin{equation}
i\frac{\partial}{\partial t}{Q}(t) = [H_+(t), Q(t)] + \{H_-(t), Q(t)\},
\label{eomQ}
\end{equation}
where $2H_\pm(t) = H(t)\pm H(t+T/2)$. In the particular case of Hamiltonian~\eqref{H},
$H_+ = \epsilon \sigma_z/2 + \beta \sigma_x$ is the time-independent part and 
$H_-(t) = \alpha \sigma_z \cos(\Omega t)$,
which appears in the anti-commutator, is the driving.

Our goal is to find an operator $Q(t)$ that complies with the 
generic properties derived so far as well as with the
specific symmetries of the model Hamiltonian \eqref{H} considered below.
In our practical calculations, the starting point will be a solution 
$\tilde Q(t)$ of Eq.~\eqref{eomQ}, which \textit{a priori} may not be unitary.
From the Hermitian adjoint of Eq.~\eqref{eomQ}, we find that it will obey
\begin{equation}
i\frac{\partial}{\partial t} \tilde Q(t) \tilde Q^\dagger(t)
= [H(t), \tilde Q(t)\tilde Q^\dagger(t)] .
\label{QdQ}
\end{equation}
This implies that if the operator $\tilde Q(t) \tilde Q^\dagger(t)$ commutes with $H(t)$,
it will be time-independent. Below we use this relation to normalize a solution 
$\tilde Q(t)$ with a time-independent factor such that it becomes unitary and still
obeys Eq.~\eqref{eomQ}.

Here, a caveat is in order. Obvious solutions of Eq.~\eqref{symmQ} read
\begin{equation}
Q(t)=\sum_{\nu} j_\nu \Pi_\nu(t),
\label{numQ}
\end{equation}
with $\Pi_\nu(t)\equiv \ketbra{\phi_\nu(t)}{\phi_\nu(t+T/2)}$. The sum encompasses a complete
set of non-equivalent Floquet modes, for instance those belonging to a given Brillouin zone,
and the coefficients $j_\nu=\pm1$ are chosen arbitrarily.  Unitarity of $Q(t)$ is
ensured by the fact that such set of modes forms a basis of the Hilbert space. However, 
without further specification, such solutions are useless for the present purpose, because 
they would assign a parity $j_\nu$  to each Floquet mode by hand. Moreover, the resulting ``symmetry''
would depend on how the Floquet modes are labeled and, hence, on the (arbitrary) choice of
the Brillouin zone. Therefore, for a meaningful parity operator, the signs $j_\nu$ must
be specified in a unique manner such that $Q(t)$ becomes a continuous function of all parameters.

\subsection{Anti-unitary symmetries}

Before attempting to solve Eq.~\eqref{eomQ}, we examine the anti-unitary
symmetries of the Floquet Hamiltonian
$H(t)-i\partial_t$ \cite{EngelhardtPRL21}.  In particular, we consider time-reversal symmetry and
particle-hole symmetry.\footnote{In accordance with Ref.~\cite{EngelhardtPRL21},
we use the terminology established for random matrices in
the context of many-body Hamiltonians \cite{AltlandPRB97}.}  The
simultaneous presence of both symmetries implies a chirality, which may substitute either one of the other two.
While these symmetries cannot explain the emergence of exact crossings,
they do facilitate the solution of Eq.~\eqref{eomQ}.

\paragraph{Time-reversal symmetry.}

Both the Hamiltonian \eqref{H} and $-i\partial_t$ possess time-reversal
symmetry as they are invariant under the transformation $\Theta =
(K,t\to-t)$, where $K$ denotes complex conjugation.  Applying $\Theta$ to
Eq.~\eqref{symmQ} and using that for all Floquet modes, $\Theta\ket{\phi_\nu(t)}$ equals
$\ket{\phi_\nu(t)}$ up to a phase factor, leads
to the operator identity
\begin{equation}
Q(t) = Q^*(-t) .
\label{trs}
\end{equation}
Therefore, the Fourier coefficients of $Q(t)$ must be real matrices.
Note that here the Pauli matrices are introduced solely for a compact
pseudospin notation.  They do not represent angular momenta and need not 
change sign under time reversal. 

\paragraph{Particle-hole~symmetry.\quad}

Transformation with $\sigma_y$ inverts the sign of Hamiltonian \eqref{H},
while complex conjugation does the same with $-i\partial_t$.  Hence, the
combined operation $C = \sigma_y K$ inverts the sign of the Floquet
Hamiltonian, i.e., $C [H(t)-i\partial_t] C^{-1} = -[H(t)-i\partial_t]$.  Therefore, the
quasienergies come in pairs with opposite sign, $q_{\bar\nu} = -q_\nu$,
and $C$ maps the corresponding Floquet modes to each other, i.e.,
$C\ket{\phi_\nu(t)}$ and $\ket{\phi_{\bar\nu}(t)}$
differ at most by a phase factor. Consequently, $Q(t) C \ket{\phi_{\nu}(t+T/2)}= j_{\bar \nu} C\ket{\phi_{\nu}(t)}$.  Acting with $C$ on 
Eq.~\eqref{symmQ} and using the relation just derived, yields
\begin{equation}
\begin{split}
C & Q(t) \ket{\phi_{\nu}(t+T/2)}
{} = j_\nu C \ket{\phi_{\nu}(t)}
\\
&{} = j_\nu j_{\bar\nu} Q(t)C\ket{\phi_{\nu}(t+T/2)} .
\end{split}
\end{equation}
Since this holds for all Floquet modes, the product
$j_\nu j_{\bar\nu}$ must be the same for all particle-hole related
pairs of modes.  Moreover, $Q(t)$ must obey the operator identity
\begin{equation}
C Q(t) C^{-1} = \sigma_y Q^*(t) \sigma_y = \mp Q(t),
\label{phs}
\end{equation}
where the upper sign holds for $j_\nu j_{\bar\nu}=-1$.

To obtain the most general form of $Q(t)$ that respects these anti-unitary
symmetries, we start with a general $2\times 2$ matrix.  Particle-hole
symmetry requires that $Q(t)$ obeys Eq.~\eqref{phs} which relates its first
and second row such that it must be of the form
\begin{align}
\tilde Q(t) & = \begin{pmatrix}
          \lambda(t) & \mu(t) \\ \pm\mu^*(t) & \mp\lambda^*(t) 
          \end{pmatrix}
\label{Qt}
\\ & \equiv \sum_k e^{-ik\Omega t}
     \begin{pmatrix}
     \lambda_k & \mu_k \\ \pm\mu_{-k} & \mp\lambda_{-k} 
     \end{pmatrix} .
\label{Qk}
\end{align}
Due to time-reversal symmetry, the Fourier coefficients of the
matrix elements, $\lambda_k$ and $\mu_k$, must be real. The tilde indicates
that so far unitarity is not ensured.

Finally, we employ Eq.~\eqref{Qd} to relate matrix elements at times
$t$ and $t+T/2$ as
\begin{align}
\label{lambdatilde}
\lambda^*(t) ={}& \lambda(t+T/2),
\\
\mu(t) ={}& \mp\mu(t+T/2),
\label{mutilde}
\end{align}
which for the Fourier coefficients means
\begin{align}
\label{lambdatildek}
\lambda_{-k} ={}& (-)^k\lambda_k,
\\
\mu_k ={}&\mp(-)^k \mu_k .
\label{mutildek}
\end{align}
While the first relation links different coefficients, the second
one yields that for the upper sign, $\mu_k=0$ when $k$ is even,
while for the lower sign, $\mu_k$ vanishes for odd $k$.

\subsection{Recurrence relations}

\begin{table*}[t]
\centering
\caption{Fourier coefficients of the functions $\lambda(t)$ and $\mu(t)$
for the integer detunings $\epsilon = n\Omega$ with $n = 0,\ldots,4$.
To achieve a compact notation, we have defined the abbreviations $D_0 = 2\Omega^2 - 2\alpha^2 + \beta^2$ and $D_1 = 6\Omega^2 - \alpha^2 - 2\beta^2$.
All coefficients with index $|k|>n$ vanish owing to the break condition.
For $n=0$, $\lambda(t)=0$ such that $Q(t) = \sigma_x$ and $J$ becomes the
generalized parity.
\label{tab:coeffs}}
\begin{tabular}{ccccccccccccc}\toprule
 & $k$ & & $-3$ & $-2$ & $-1$ & $0$ & 1 & 2 & 3 & \quad4\quad
\\ \midrule
$n=0$ & $\lambda_k$ &&&&& -
\\
 & $\mu_k$ &&&&& $1$ &
\\ \midrule
$n=1$ & $\lambda_k$ &&&& - & $\beta$ & -
\\
 & $\mu_k$ &&&& - & - & $\alpha$
\\ \midrule
$n=2$ & $\lambda_k$ &&& - & $-\alpha\beta$ & $\beta\Omega$ & $\alpha\beta$ & -
\\
  & $\mu_k$ &&& - & - & $\beta^2$ & - & $\alpha^2$
\\ \midrule
$n=3$ & $\lambda_k$ && - & $\alpha^2\beta$ & $-2\alpha\beta\Omega$ &
$\beta(2\Omega^2-\alpha^2+\beta^2)$ &
$2\alpha\beta\Omega$ & $\alpha^2\beta$ & -
\\
  & $\mu_k$ && - & - & $-\alpha\beta^2$ & - & $2\alpha\beta^2$ & - & $\alpha^3$
\\ \midrule
$n=4$ & $\lambda_k$ && $-\alpha^3\beta$ & $3\alpha^2\beta\Omega$ &
$-\alpha\beta D_1$ &
$2\beta\Omega D_0$ & $\alpha\beta  D_1$ &
$3\alpha^2\beta\Omega$ & $\alpha^3\beta$ & -
\\
  & $\mu_k$ && - & $\alpha^2\beta^2$ & - &
$\beta^2 D_0$ & - & $3\alpha^2\beta^2$ & - & $\alpha^4$
\\
\bottomrule
\end{tabular}
\end{table*}

Inserting the Fourier series \eqref{Qk} into the equation of motion~\eqref{eomQ} yields a
set of four coupled recurrence equations for the matrix elements of the 
coefficients $\tilde Q_k$, where only two equations are independent.  
From the diagonal matrix elements, one finds
\begin{equation}
-k\Omega\lambda_k - \beta(\mu_k\mp\mu_{-k}) +
\alpha(\lambda_{k-1}+\lambda_{k+1}) = 0 .
\label{recurrence-diag}
\end{equation}
The off-diagonal matrix elements provide the relation
\begin{equation}
\begin{split}
(\epsilon-k\Omega)\mu_k
={} & \beta(\lambda_k\pm\lambda_{-k})
\\
={} & \beta[1\pm(-)^k]\lambda_k ,
\label{recurrence-offdiag}
\end{split}
\end{equation}
where the second equality follows from Eq.~\eqref{lambdatildek}.  This
relation contains only Fourier coefficients with equal index and can be
used to eliminate in Eq.~\eqref{recurrence-diag} the dependence on
$\mu_{\pm k}$.

Once more, we make use of knowledge of the CDT case $\epsilon=0$ in which
$Q(t)=\sigma_x$ consists of only the Fourier component with $k=0$.  While
this appears impossible in the detuned case, we employ a slightly weaker
condition, namely that the symmetry operator should have a finite number of
Fourier components.  Let us therefore assume that there exists an integer $n\geq 0$ such
that $\tilde Q_k=0$ and, thus, $\lambda_k = 0 = \mu_k$ for all $|k|>n$.  In
particular, $\mu_{n+1} = \lambda_{n+1} = \lambda_{n+2} = 0$, such that
Eq.~\eqref{recurrence-diag} for $k=n+1$ reads
\begin{equation}
\lambda_n = 0 ,
\label{recurrence_lambda_n}
\end{equation}
while Eq.~\eqref{recurrence-offdiag} simplifies to
\begin{equation}
(\epsilon - n\Omega)\mu_n = 0 .
\label{recurrence_mu_n}
\end{equation}
A non-trivial solution requires that the
prefactor $\epsilon-n\Omega$ in the latter equation vanishes, i.e., the
detuning must match an integer multiple of the driving frequency, $\epsilon =
n\Omega$.  Such integer detuning has been identified as condition for the
existence of a hidden symmetry also for the Rabi model
\cite{ReyesBustosJPA21}.  Owing to the assumption $\epsilon>0$, Eq.~\eqref{recurrence_mu_n} also 
implies $\mu_{-n} = 0$, such that $\mu_n$ remains the only
non-vanishing matrix element of $\tilde Q_{n}$.  This finally allows us to set the
sign of $Q(t)$ in a unique manner and independently of the parameter values 
by choosing $\mu_n$ real and positive.  For convenience, we set 
$\mu_n = \alpha^n$ and take care for a proper normalization later.  Together 
with Eq.~\eqref{mutildek}, this determines a further sign,
namely the one in Eq.~\eqref{phs}, which must read $\mp = (-)^{n+1}$.

Summarizing the relations obtained so far, we find that for $|k|<n$,
the recurrence relation in Eq.~\eqref{recurrence-diag} can be written as
\begin{equation}
\alpha\lambda_{k-1} - (k\Omega+b_k)\lambda_k + \alpha\lambda_{k+1} = 0 ,
\label{recurrence_lambda}
\end{equation}
where the second term has been expressed in terms of $\lambda_k$ by making
use of Eq.~\eqref{recurrence-offdiag} and introducing the shorthand notation
\begin{equation}
b_k =
\begin{cases}
\frac{4k\beta^2}{(n^2-k^2)\Omega} & \text{if $n+k$ even},
\\
0 & \text{else}.
\end{cases}
\end{equation}
With the boundary condition $\lambda_n = \lambda_{n+1}=0$, its evaluation
is straightforward.  Since $b_{-k} = -b_k$, the solution will be consistent with Eq.~\eqref{lambdatildek} and with the condition that all $\lambda_k$ must vanish for
$k\leq -n$.

In a last step, we have to normalize $Q(t)$ such that it becomes unitary and still obeys
the equation of motion \eqref{eomQ}.
This would be impossible if the required normalization factor were time-dependent.
Here, however, the form of $\tilde Q(t)$ in Eq.~\eqref{Qt}
ensures that $\tilde Q(t) \tilde Q^\dagger(t)$ is proportional to a unit matrix.  Then
the right-hand side of Eq.~\eqref{QdQ} vanishes, and we can conclude that
$\tilde Q(t) \tilde Q^\dagger(t)$ is time-independent.  Therefore, any 
solution of Eq.~\eqref{eomQ} that complies with particle-hole symmetry can be
normalized by a time-independent factor such that the corresponding unitary $Q(t)$ obeys 
the same equation of motion.

The resulting Fourier coefficients $\lambda_k$ and $\mu_k$ for integer detuning up to $\epsilon =
4\Omega$ are compiled in Table~\ref{tab:coeffs}.  In the time domain, for
$n=1$,
\begin{equation}
Q(t) \propto \begin{pmatrix} \beta & \alpha e^{-i\Omega t} \\
-\alpha e^{i\Omega t} & \beta \end{pmatrix}
\label{Q1}
\end{equation}
while for $n=2$,
\begin{align}
&Q(t)
\\ \nonumber & \propto
\begin{pmatrix} \beta\Omega-2i\alpha\beta \sin(\Omega t) & \beta^2+\alpha^2 e^{-2i\Omega t} \\
\beta^2+\alpha^2 e^{2i\Omega t} & -\beta\Omega-2i\alpha\beta \sin(\Omega t)
\end{pmatrix} .
\label{Q2}
\end{align}
These operators become unitary upon division by the square roots of
$\alpha^2+\beta^2$ and $(\alpha^2+\beta^2)^2+(\beta\Omega)^2$,
respectively.  For $n=1$, an intuitive alternative derivation of $Q(t)$ is
provided in Appendix \ref{app:n=1}.

\subsection{Consequences for the Floquet spectrum}

Since $J(t)$ is $T$-periodic, it can be considered as an operator
in Sambe space, just like the Floquet Hamiltonian $H(t)-i\partial_t$.
Then, from Eq.~\eqref{Qd}, it follows that it is Hermitian.
Equations \eqref{Feq} and \eqref{symmJ} identify the Floquet modes as
their common eigenstates.  Hence, each mode $\ket{\phi_\nu(t)}$ can be
characterized by the corresponding eigenvalues, namely a quasienergy
$q_\nu$ and a parity $j_\nu$.  The latter is given by the expectation
value
\begin{equation}
j_\nu
= \frac{1}{T} \int_0^T dt\, \bra{\phi_\nu(t)} J(t) \ket{\phi_\nu(t)}
= \pm 1 ,
\label{jsambe}
\end{equation}
with the time integration stemming from the inner product in Sambe space
\cite{SambePRA73}.  The practical computation is simplified by the fact that
Eq.~\eqref{symmQ} holds already in Hilbert space, such that
\begin{equation}
j_\nu = \bra{\phi_\nu(t)} Q(t) \ket{\phi_\nu(t+T/2)} ,
\label{j}
\end{equation}
which turns out to be time independent.  In our numerical calculations, we
evaluate it at $t=0$.

Generally the eigenvalues of Hermitian operators, as a function of any
parameter, exhibit level repulsion, unless they belong to modes
from different symmetry classes \cite{Haake2018}.  In the present case,
such symmetry is given by the hidden parity $J(t)$, which, for
integer detuning, allows the emergence of exact crossings.
To visualize this behavior, we have diagonalized numerically for
$\epsilon=\Omega$ the Floquet Hamiltonian to obtain the quasienergies and
the Floquet modes, as well as the expectation value of $J(t)$ determined by
Eq.~\eqref{j} with the (normalized) operator $Q(t)$ in Eq.~\eqref{Q1}.  
Figure~\ref{fig:spectrum} shows the resulting Floquet spectrum extended 
over three Brillouin zones.  The color of the lines reflects the value 
of $j_\nu$. The avoided and exact crossings verify the consequences of the 
hidden parity.

This observation is consistent with the von Neumann-Wigner theorem
\cite{Haake2018} which states that in the presence of time-reversal
symmetry, one needs to adjust two independent parameters to obtain a
degeneracy.  Here, this is done in the following way.  For arbitrary values
of $\Omega$ and $\beta$, one has to (i) adjust the detuning to an integer
multiple of the driving frequency and (ii) choose a particular amplitude
$\alpha$.  Alternatively, one may start with an arbitrary value of $\alpha$
and will find degeneracies for particular values of $\beta$.  It is
elucidating to consider non-integer detunings in the context of the von
Neumann-Wigner theorem.  For arbitrary fixed detuning $\epsilon$, one may search for values of
$\alpha$ and $\beta$ such that the quasienergies are degenerate.  However,
there seems to exist only the trivial solution $\beta=0$.  For this value,
however, the nature of the problem is entirely different, because
the system possesses the time-local symmetry $\sigma_z$.

\section{Numerical computation of the symmetry operator}
\label{sec:numerics}

\begin{figure*}
\centerline{\includegraphics{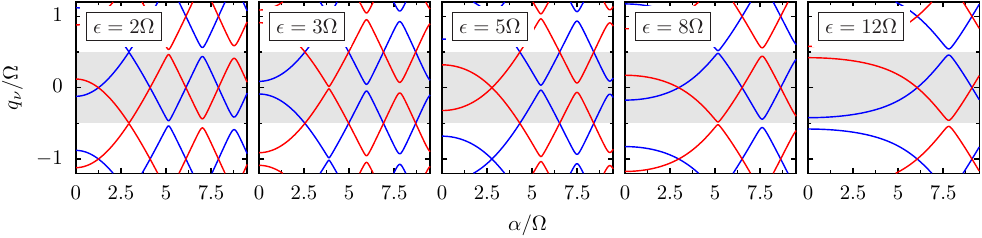}}
\caption{Floquet spectra as a function of the driving amplitude for various
integer detunings $\epsilon = n\Omega$.  The tunnel matrix element $\beta =
2.7\Omega$ and the color code of the parity are as in
Fig.~\ref{fig:spectrum}.  As there, some crossings of levels with equal
parity appear exact, but in fact are narrowly avoided.}
\label{fig:spectra}
\end{figure*}

While we have already proven that for any integer detuning $\epsilon =
n\Omega$, a hidden parity exists, its analytical calculation becomes
increasingly tedious for larger $n$.  For a numerical solution, one may
follow the scheme derived above and numerically iterate the recurrence relation
\eqref{recurrence_lambda}.  This will provide the Fourier coefficients
$\lambda_k$ and $\mu_k$, and eventually $Q(t)$.
Here, however, we follow a different route, because it is instructive
to obtain the Fourier coefficients $Q_k$ directly from the Floquet modes.  
Moreover, such numerical solution is not restricted to the two-level system 
and, thus, hints on how to compute time-nonlocal symmetries for other systems.

To this end, we start from Eq.~\eqref{numQ} and write the Fourier components of $Q(t)$ in the form
\begin{equation}
Q_k = \sum_\nu j_\nu \Pi_{\nu,k},
\label{numQ1}
\end{equation}
where the operators $\Pi_{\nu,k}$ are the Fourier components of $\Pi_\nu(t)$ defined
in Sec.~\ref{sec:LRI}. These can be expressed in terms of the sidebands of the Floquet modes as
$\Pi_{\nu,k} = \sum_{k'} (-1)^{k'} \ket{\phi_{\nu,k'+k}}\bra{\phi_{\nu,k'}}$.
The sum is restricted to $d$ non-equivalent Floquet modes, 
where $d$ is the dimension of the Hilbert space. While the $\Pi_{\nu,k}$ depend on the
choice of the Brillouin zone, it is crucial that the resulting operator $Q(t)$ is independent
of this arbitrariness.

The analytical insight gained so far suggests the need for a break condition. We therefore
assume that the Fourier coefficients $Q_k$ in Eq.~\eqref{numQ1} vanish when $|k|$
exceeds some \textit{a priori} unknown value. This assumption leads to an overdetermined
set of homogeneous linear equations for the coefficients $j_\nu$, which in general admits
only the trivial solution $j_\nu=0$. A non-trivial solution for the $j_\nu$ can then exist
only if a time-nonlocal symmetry is present. In that case, such a solution can be obtained
from any subset of $d$ equations with sufficiently large Fourier index, while the remaining
equations may be used for consistency checks. Since the system of equations is homogeneous,
the solution is determined only up to an overall sign (or, more generally, a phase factor).
As in the analytical construction, this residual freedom must be fixed in a unique manner.
Based on the considerations above, this can be achieved by choosing the only non-vanishing
matrix element of $Q_k$ with the largest index to be real and positive.
For details of the implementation of the numerical scheme, see Appendix~\ref{app:numerics}.

To verify the existence of a hidden time-nonlocal symmetry also for
$\epsilon=n\Omega$ with $n>1$, we have computed the Floquet modes for
various integer detunings by diagonalizing the Floquet Hamiltonian.  To
find $Q(t)$, we have computed the $\Pi_{\nu,k}$ and have determined the
parities $j_\nu$ as described in the last paragraph.  Then for each Floquet mode, we have
evaluated Eq.~\eqref{j} at time $t=0$ to obtain the parity of each mode,
including all equivalent ones.  The corresponding expression in Sambe
space, Eq.~\eqref{jsambe}, has been used for confirmation.
Figure~\ref{fig:spectra} shows the resulting spectra, where again the color
of the curves refers to the value of $j_\nu$.  Within numerical precision it
assumes the values $\pm 1$.  Besides verifying our conjecture for the 
crossings, this also demonstrates that our numerical approach is reliable even 
for rather large integer detunings.

Finally, let us us remark that the numerical scheme with the time-local ansatz
$J(t) = \sum_{\nu} j_\nu |\phi_\nu(t)\rangle\langle\phi_\nu(t)|$
leads to the trivial result $J(t)=\mathbb{I}$.  This underlines that time-nonlocality is an
essential constituent of the present symmetry.

\section{Discussion and conclusions}
\label{sec:conclusions}

We have developed the concept of hidden time-nonlocal Floquet symmetries.
It is based on an ansatz with a spatio-temporal transformation that
resembles the classic generalized parity \cite{PeresPRL91} known from CDT.
The main difference is that the spatial part now is time dependent.  Its
properties can be determined from a conjectured automorphism of the Floquet
modes.  It turned out that the mapping $Q(t)$ must reflect all symmetries of
the Hamiltonian, foremost the time periodicity.
The practical calculation requires solving a Liouville-like equation in which
the driving Hamiltonian appears within an anti-commutator.

For the driven two-level system with integer detuning, we have demonstrated
the existence of such symmetry.  The constructive proof makes use of the
anti-unitary symmetries of the Hamiltonian, while an explicit expression
can be found from a recurrence equation together with a break condition.
The hidden parity partitions the Sambe space into even and odd subspaces, allowing quasienergies from different subspaces to form exact crossings.
We have verified numerically that the Floquet spectrum of our
model exhibits this feature.
The condition of integer detuning has been found also for the existence of a hidden integral
of motion of the Rabi model \cite{AshhabPRA20, MangazeevJPA21, ReyesBustosJPA21}.
Despite this similarity, we note several remarkable differences, such as quasienergies not being bounded from below, the possibility of exploiting anti-unitary symmetries, and the availability of a constructive existence proof.

For the numerical computation of the symmetry operator, we have developed
an independent scheme not based on the recurrence relation.  It turned out
rather stable even for large detunings.  Besides enabling a visualization in 
the two-level case, it provides a tool for the search for hidden time-nonlocal
symmetries in other Floquet systems.

\begin{acknowledgments}
This work was supported by the Spanish Ministry of Science, Innovation, and Universities under Grant No.\ PID2023-149072NB-I00 (S.K.); by MCIN/AEI/10.13039/501100011033 and by ``ERDF A way of making Europe'', EU, under Grant No.\ PID2022-136228NB-C22 (J.C.-P.); and by the CSIC Quantum Technologies Platform QTEP (S.K.). It was also co-financed by the European Union, the Ministerio de Hacienda y Funci\'on P\'ublica, FEDER, and the Junta de Andaluc\'{\i}a under Project SOL2024-31833 (J.C.-P.).

\end{acknowledgments}

\appendix

\section{Intuitive solution for \texorpdfstring{$n=1$}{n=1}}
\label{app:n=1}

For the smallest non-trivial integer detuning, $\epsilon=\Omega$, the
invariant can be obtained in a less formal way.  The idea is to find a
sequence of transformations that inverts the sign of the driving, which 
finally is canceled by the time shift $P$.  The mapping starts with a
transformation to the interaction picture with respect to the detuning,
$U_0(t) = \exp(-i\sigma_z\Omega t/2)$, which results in
\begin{equation}
\tilde H (t) = \beta\sigma_x\cos(\Omega t) - \beta\sigma_y\sin(\Omega t)
+ \alpha\sigma_z\cos(\Omega t) .
\end{equation}
Then transformation with $\beta\sigma_x + \alpha\sigma_z$ (we ignore
the normalization) flips the sign of the second term of $\tilde H (t)$, while
leaving the rest as is.  A further transformation with $U_0(t)$ yields
\begin{equation}
H'(t) = -\frac{\Omega}{2}\sigma_z + \beta\sigma_x
+ \alpha\sigma_z\cos(\Omega t) ,
\end{equation}
which is the original Hamiltonian, but with inverted detuning.  Finally,
transformation with $\sigma_x$ moves the minus sign to the time-dependent
term, such that the full transformation
\begin{equation}
\tilde Q(t) = U_0(t) (\beta\sigma_x + \alpha\sigma_z) U_0(t) \sigma_x
\end{equation}
agrees with Eq.~\eqref{Q1}.

\section{Numerical solution: Implementation}
\label{app:numerics}

It is instructive to discuss in more detail the implementation of the numerical approach outlined in 
Sec.~\ref{sec:numerics}. We seek an operator $Q(t)$ of the form \eqref{numQ} with a finite set of Fourier 
coefficients $Q_k$. With regard to parity symmetry, the coefficients $j_\nu$ are restricted to $\pm 1$.

The numerical computation is based on the eigenvalue equation~\eqref{Feq}, 
which provides the Floquet modes and, consequently, the operators $\Pi_\nu(t) 
= \ket{\phi_\nu(t)}\bra{\phi_\nu(t+T/2)}$.  To bring Eq.~\eqref{numQ} into a form
suitable for a numerical implementation, we rearrange
the Fourier coefficients of the $2\times 2$ matrices $\Pi_{\nu}(t)$ and $Q(t)$ to obtain the
4-dimensional column vectors $\mathbf\Pi_{\nu,k}$ and
$\mathbf Q_k$.  Then the ansatz takes the form
\begin{equation}
\begin{pmatrix}
\vdots & \vdots \\
\mathbf\Pi_{1,k_0} & \mathbf\Pi_{2,k_0} \\
\vdots & \vdots \\
\mathbf\Pi_{1,1} & \mathbf\Pi_{2,1} \\
\mathbf\Pi_{1,0} & \mathbf\Pi_{2,0} \\
\mathbf\Pi_{1,-1} & \mathbf\Pi_{2,-1} \\
\vdots & \vdots \\
\mathbf\Pi_{1,-k_0} & \mathbf\Pi_{2,-k_0} \\
\vdots & \vdots
\end{pmatrix}
\begin{pmatrix} j_1 \\ j_2 \end{pmatrix}
=
\begin{pmatrix}
\vdots \\ \mathbf 0 {\vphantom{\Pi_{1,k_0}}} \\ \vdots \\ \mathbf Q_{1} \\
\mathbf Q_0 \\ \mathbf Q_{-1} \\ \vdots \\ \mathbf 0 {\vphantom{\Pi_{1,k_0}}} \\ \vdots
\end{pmatrix} ,
\label{LSE}
\end{equation}
which determines the coefficients $Q_k$ up to a factor.
A key non-trivial ingredient is the truncation condition $\mathbf Q_k = \mathbf 0$ for $|k| \geq k_0$.
For our model, we already know that for $k_0$, we may choose any integer $k_0 > n$.
Without this knowledge, one needs a tradeoff between exceeding the index of the
last non-vanishing $\mathbf Q_k$ and avoiding the numerical uncertainties of the
Fourier components with very large index.

For the numerical implementation of Eq.~\eqref{LSE}, we choose for the Fourier 
components a cutoff $K>k_0$, such that the matrix on the left-hand side has dimension $4(2K+1)\times 2$. 
Then any two rows of the resulting matrix-vector equation with zero on the right-hand
side constitute a set of two homogeneous linear equations,
which generally admits only the trivial solution $j_1=j_2=0$.  However,
if a symmetry with the desired properties exists, one finds a non-trivial solution for $j_\nu$, which can be
normalized such that $j_1=1$ and $j_2=\pm1$.  This solution must be consistent with all other rows of 
Eq.~\eqref{LSE} with $|k| \geq k_0$.  The central rows then yield the non-vanishing coefficients $Q_k$ and, 
thus, $Q(t)$.  However, owing to the arbitrary normalization $j_1=1$, the symmetry operator $Q(t)$ is 
determined only up to a sign, which must be fixed uniquely in the final step.  In the 
present  case, this is straightforward, because the last non-vanishing coefficient $Q_n$
has only one non-zero matrix element, which can be chosen real and positive.

\end{document}